\documentclass[
    aps,
    pra,
    twocolumn,
    superscriptaddress,
    nofootinbib,
    longbibliography
]{revtex4-2}

\usepackage{amsmath}
\usepackage{amssymb}
\usepackage{bm}
\usepackage{graphicx}
\usepackage{physics}
\usepackage{siunitx}
\usepackage{xcolor}
\usepackage{booktabs}
\usepackage{microtype}
\usepackage{xpatch}
\usepackage{tikz}
\usepackage{gensymb}
\usepackage[caption=false]{subfig}
\usepackage[
    colorlinks=true,
    linkcolor=blue,
    citecolor=blue,
    urlcolor=blue
]{hyperref}
\usepackage{appendix}

\newcommand{\equalcontribnote}{%
    $^{\dagger}$These authors contributed equally to this work.%
}

\makeatletter

\xpatchcmd{\titleblock@produce}
    {\frontmatter@author@produce@script}
    {%
        \frontmatter@author@produce@script
        \par\addvspace{4\p@}%
        {%
            \centering\small\itshape
            \equalcontribnote
            \par
        }%
    }
    {\class@info{Equal-contribution note patched in.}}
    {\class@warning{Could not patch the equal-contribution
        note into the title block.}}

\renewcommand{\@fnsymbol}[1]{%
    \ensuremath{%
        \ifcase#1\or
            *\or
            \ddagger\or
            \mathsection\or
            \mathparagraph\or
            **\or
            \ddagger\ddagger\else
            \@ctrerr
        \fi
    }%
}

\makeatother

\begin{document}

\title{{High Resolution Hyperspectral Quantum Infrared Imaging with Undetected Photons}}

\author{Aadi Suketu Bharatia$^{\dagger}$}
\affiliation{%
    Department of Physics,
    Indian Institute of Technology Bombay,
    Mumbai--400076, India
}

\author{Vidya Prakash Muppalla$^{\dagger}$}
\affiliation{%
    Department of Physics,
    Indian Institute of Technology Bombay,
    Mumbai--400076, India
}

\author{Ananya Asthana}
\affiliation{%
    Department of Biosciences and Bioengineering,
    Indian Institute of Technology Bombay,
    Mumbai--400076, India
}

\author{Priyanka Kandankel}
\affiliation{%
    Department of Physics,
    Indian Institute of Technology Bombay,
    Mumbai--400076, India
}

\author{Tanmoy Chakraborty}
\email{tanmoy@rri.res.in}
\affiliation{%
    Raman Research Institute,
    Bengaluru--560080, India
}

\author{Sreelaja Nair}
\email{sreelaja@iitb.ac.in}
\affiliation{%
    Department of Biosciences and Bioengineering,
    Indian Institute of Technology Bombay,
    Mumbai--400076, India
}

\author{Anshuman Kumar}
\email{anshuman.kumar@iitb.ac.in}
\affiliation{%
    Laboratory of Optics of Quantum Materials (LOQM),
    Department of Physics,
    Indian Institute of Technology Bombay,
    Mumbai--400076, India
}
\affiliation{Centre of Excellence in Quantum Information, Computation, Science and Technology, Indian Institute of Technology Bombay, Powai, Mumbai- 400076, India}

\begin{abstract}

Quantum sensing in the infrared regime often demands the use of expensive and cryogenically operated detectors. Nonlinear interferometry has paved the way for information transfer from the infrared to the near--infrared or visible regime in various quantum technologies. A particularly interesting application of these interferometers involves spectrally--resolved quantum imaging using undetected photons, enabled by the use of temperature--tunable parametric amplifiers. In this work, we demonstrate {spectrally resolved} transmission imaging with undetected photons around 1550 nm using {a} spontaneous parametric down conversion based folded nonlinear interferometric scheme, registering the attenuation of a zebrafish larva as a biological feasibility test. We characterize and calibrate the interferometer, having achieved spectrally resolved interference fringes with a maximum baseline visibility of $\sim 74\%$ in the near--infrared regime. {Phase--stepped readout of a grating spectrometer gives a resolving power of $\sim1460$ (from the spectrometer's point spread function)  at 1550~nm, and $\sim 730$ from the coherence length, the finest reported for imaging with undetected photons, over an instantaneous spectral window of $\approx 51$~cm$^{-1}$.}

\end{abstract}


\maketitle

\section{Introduction}

Most optical information that distinguishes one material from another or one biological tissue from another is often contained in the infrared regime \cite{EvansGeo}. In this regime, molecular bonds absorb through their fundamental vibrations, so a measurement of the infrared transmission coefficient reports directly, the chemical composition of the sample. Infrared signals also travel through biological tissues farther than visible light \cite{Smith2009SecondWF,hong2017nir,TaggJesus}, the primary hindrance to exploiting which is detection hardware: infrared detection relies on InGaAs, InSb, or MCT arrays, which are far more expensive than silicon detectors, offer far fewer pixels, and require cryogenic cooling to keep dark--counts under control.  In comparison, silicon cameras are inexpensive, quiet, megapixel--scale, and readily accessible. 

Non-linear interferometry provides a convenient platform to achieve infrared probing and silicon--based detection. Zou, Wang and Mandel showed in 1991, that if two parametric down--conversion sources are pumped coherently, aligning the idler beam of the first with that of the second leads to the interference of the corresponding signal beams, which vanishes when the idler photon from the first source is blocked, even though no idler photon is actually detected \cite{zouWangMandel,wangZouMandel}. The coherence of the signal photons is induced by the indistinguishability of the corresponding idler photons, without any stimulated emission taking place. Such an arrangement represents a nonlinear, or SU(1,1) type of interferometer \cite{yurke1986,Chekhova:16,HochrainerZeilinger}. Lemos \textit{et al.} turned this into an imaging method by recording the silhouette of an object placed in a 1550 nm idler beam while detecting only 810 nm photons \cite{LemosZeilinger}. Such an SU(1,1) interferometer--based quantum imaging technique now has a detailed theoretical treatment given in \cite{LahiriZeilinger,BarretoLemos:22}. It is worth noting that Induced Coherence (IC) has a classical analogue under strong seeding \cite{Shapiro2015}. Thus, the quantum character of such an experiment depends on operating at very low parametric gain, where at most one photon pair is generated from a single pump photon.

\begin{figure*}[ht]
  \centering
  \begin{tikzpicture}
    \node[anchor=south west, inner sep=0] (main)
      {\includegraphics[width=\textwidth]{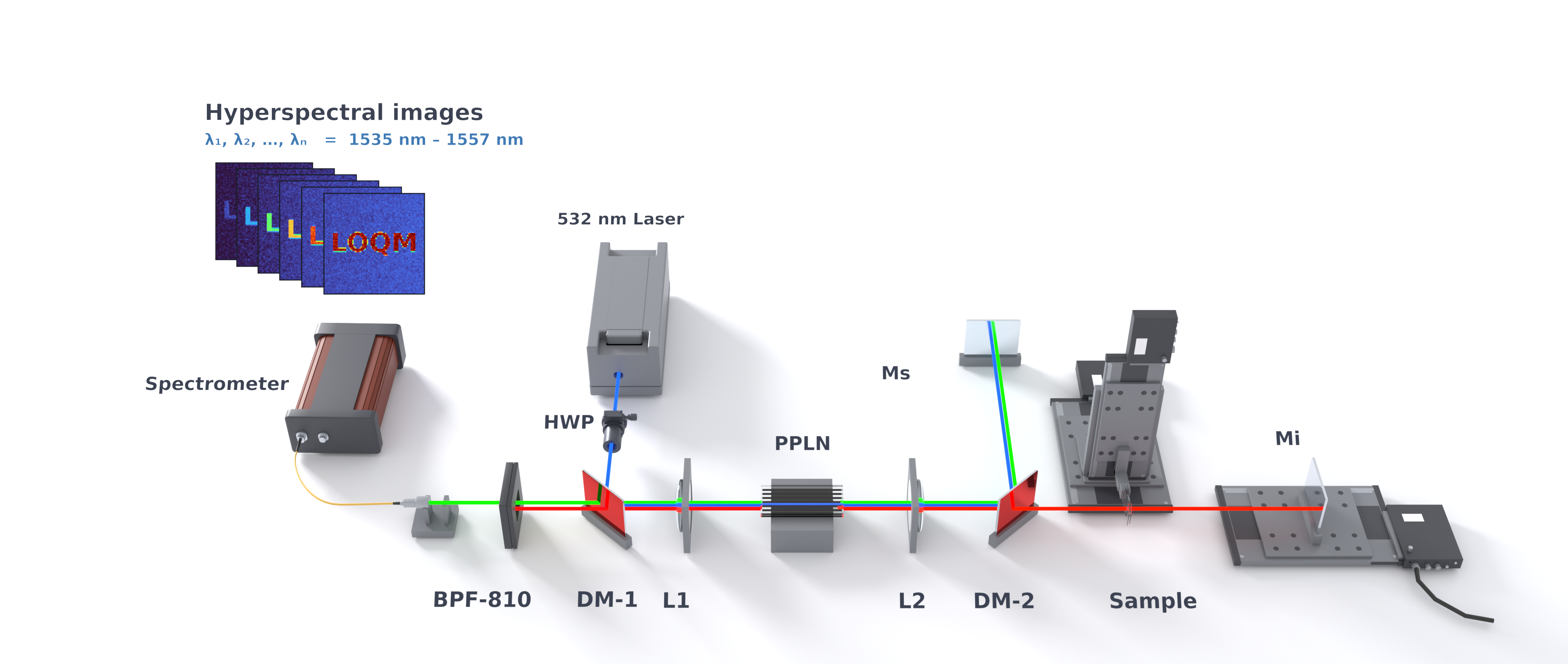}};

    \node[anchor=north east, draw=black, line width=0.4pt,
          fill=white, inner sep=1pt,
          xshift=-4pt, yshift=-4pt] at (main.north east)
      {\includegraphics[width=0.25\textwidth]{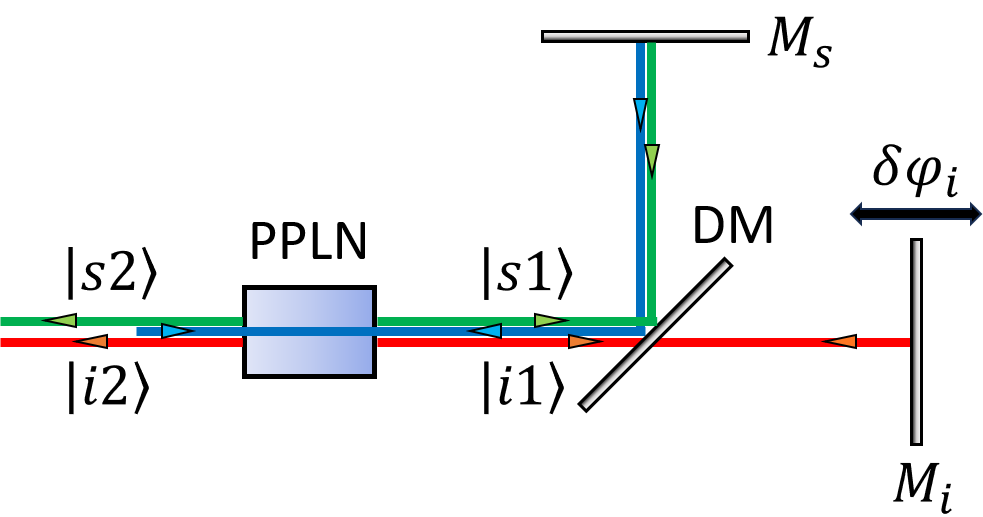}};
  \end{tikzpicture}
  \caption{Experimental setup: blue: pump beam, green: signal beam, red: idler beam, DM: dichroic mirror, L: lens, BPF: band--pass filter, $M_s$: signal mirror, $M_i$: idler mirror}
  \label{fig:wide_with_inset}
\end{figure*}

In a little more than a decade, this principle has produced different applications. Absorption and refractive index spectroscopy of solids and gases have been performed \cite{kalashnikov2016, paterova2018infrared,Lindner_2020}. Wide-field microscopy has been pushed to the molecular fingerprint region, with images across 3.4 - 4.3 $\mu\text{m}$ at a 35 $\mu\text{m}$ resolution recorded on a silicon sCMOS camera \cite{kviatkovskyRamelow}. Further, the same principle has been applied to non--destructive inspection of silicon devices \cite{PaterovaSemiconductors}. Two results bear most directly on the present work: Paterova \textit{et al.} demonstrated wide--field hyperspectral infrared microscopy by dispersing the signal photons from a crystal superlattice, mapping the chemical contrast of a patterned test sample \cite{paterova2020hyperspectral,toa2021broadbanddiffractioncorrelatedphotons}. { More recently, Placke \textit{et al.} combined a wide--field nonlinear interferometer with pixel-wise Fourier-transform spectroscopy to cover the 2300- 3100 $\text{cm}^{-1}$ wavenumber regime across 3500 spatial modes \cite{placke2025midirhyperspectralimagingundetected}}. {Gattinger \textit{et al.} have since read a static interferometer with a grating spectrometer for scanless hyperspectral mapping, reaching 8~cm$^{-1}$ \cite{gattinger2026scanless}. In single--point spectroscopy, Fourier--transform readout reaches 0.5--6~cm$^{-1}$ \cite{Lindner_2020,kaur2024pulsed}, and phase--scanned readout of a grating spectrometer reaches the instrument--limited 1.5~cm$^{-1}$ \cite{kaufmann2022grating}; Kaufmann \textit{et al.} proposed the latter as a route to hyperspectral imaging. Camera--based imaging on the same 532~nm--pumped PPLN platform at 808/1558~nm has been demonstrated with band--pass filtering, without spectral resolution \cite{Pearce_2023}.}

Conventional wide--field microscopy collects out--of--focus background, while confocal rejects it at the cost of fewer detected photons per pixel, forcing brighter illumination or longer dwell times; both accumulate phototoxicity over an imaging session \cite{icha2017phototoxicity}. Imaging with undetected photons sidesteps this. The photons that probe the sample are never detected, allowing for label--free measurements at ultra--low power \cite{GilaberteBasset2021VideoRateIW}. We propose to use zebrafish larvae as a biological sample, as they share a $70\%$ genetic similarity with humans and are widely used in biomedical and fundamental biology research. In the infrared region around 1550 nm, scattering in biological tissues is reduced, and absorption by chromophores like melanin in pigment cells and haemoglobin is relatively lower. Also, zebrafish larvae have been shown to demonstrate increased pigment transparency in this region, extending the developmental window for effective optical imaging \cite{volkov2022swir}.

In this work, we report {spectrally resolved} transmission imaging with undetected photons around 1550 nm, using a folded single--crystal nonlinear interferometer, and apply it to an intact zebrafish larva as a biological feasibility test. {We read the interferometer out by stepping the idler phase and recording every spectrometer channel independently, which transfers the full spectrometer resolution to the idler.} We characterise the instrument by its spectrally resolved baseline visibility, which peaks at $74.4\%$ near 811 nm, and verify that the measured fringe period follows the energy-conservation prediction across the band. We then raster--scan two samples: a laser--printed carbon--ink target and a fixed zebrafish larva, and reconstruct, at each idler wavelength, a map of the transmission inferred from the interferometric fringe visibility. 

An upper bound of $\mu < {3.7}\times10^{-6}$ photons per electromagnetic mode places our experiment firmly in the single-pair regime (Appendix E). To the best of our knowledge, this is the first spectrally resolved measurement with undetected photons performed on zebrafish in the short--wave infrared region. We present it as a feasibility demonstration: the spatial resolution is currently set by the $\sim$ 1 mm FWHM idler beam waist (computed by measuring the signal beam waist on an sCMOS camera). The one--to--one correspondence between signal and idler spectral modes means each detected signal mode reports on a single conjugate idler mode \cite{Fuenzalida_2022}. Spectral resolution of the collected signal beam therefore rejects photons whose idler partners missed the probed region, suppressing background and crosstalk between neighbouring wavelengths.

\section{Theory}
In the following section, we briefly discuss Induced Coherence in nonlinear interferometers \cite{wangZouMandel, Chekhova:16}, with a focus on the expected theoretical signal interferogram, mapped to the object's optical transmission at the idler wavelength. Further, we elaborate on the setup's wavelength--tunable architecture, enabled by the PPLN crystal.

\subsection{Induced Coherence}

We observe interference in the signal photons generated from two SPDC processes, made indistinguishable by mode--matching their corresponding idler photons \cite{wangZouMandel}. Thus, indistinguishability of the idler photons generated from these processes induces coherence in their signal counterparts. Several schemes have been proposed to realize this effect experimentally \cite{chakraborty2025magneto, paterova2018infrared, paterova2019polarization}. We use a `folded' configuration \cite{paterova2018infrared, paterova2019polarization}, wherein both nonlinear processes occur in the same crystal bidirectionally (Figure \ref{fig:wide_with_inset} inset), as elucidated in section III. Thus, the total wavefunction of the system after the second pass through the crystal, as mentioned in \cite{paterova2019polarization} is given by
\begin{equation}
|\Psi\rangle = |0\rangle + \eta \hat{a}_{s1}^\dagger \hat{a}_{i1}^\dagger |0\rangle + \eta \hat{a}_{s2}^\dagger \hat{a}_{i2}^\dagger |0\rangle
\end{equation}
where $|0\rangle$ denotes the vacuum state, $\eta$ denotes the SPDC efficiency, and $(s1(2),i1(2))$ denotes the signal--idler photon pair generated after the first (second) pass through the crystal (Figure \ref{fig:wide_with_inset} inset). It should be noted that Eq. (1) assumes that the signal and idler photons are generated into single--modes. 

The extent of indistinguishability between the idler modes indexed by $i1$ and $i2$ can be tuned by placing an object after the first pass. For an object with a complex transmission coefficient $\tau$, the photon annihilation operator for mode $i2$ can be approximately written as 
\begin{equation}
\hat{a}_{i2} \approx e^{i\varphi_i}\tau^2\hat{a}_{i1} + \hat{\ell}
\end{equation}
where $\varphi_i$ denotes the acquired idler phase after the first pass and $\hat{\ell}$ accounts for lossy modes \cite{paterova2019polarization}. The count--rate for signal photons is given by $R_s \propto \langle\Psi|\hat{E}_s^{(-)}\hat{E}_s^{(+)}|\Psi\rangle$ \cite{paterova2019polarization}, which yields the following expression \cite{paterova2019polarization}: 
\begin{equation}
R_s \propto 1 + |\tau|^2|\mu(\Delta t)|\cos\left(\Phi -\varphi_i + \arg\tau^2\mu(\Delta t)\right)
\end{equation}
where $\Phi$ accounts for the pump and signal phases \cite{Heuer}, and $\mu(\Delta t)$ is the normalized SPDC correlation function for a time delay $\Delta t$ between the arrival times of signal photons from the two SPDC events on the detector, given by \cite{paterova2019polarization} (see Appendix A for details)
\begin{equation}
\mu(\Delta t) = \int |F(\Omega)|^2 e^{-i\Omega \Delta t}\ d\Omega
\end{equation}
where $F(\Omega)$ is the normalized bi-photon SPDC amplitude at a frequency de--tuning of $\Omega$ from the central signal and idler frequencies. Our use of this interferometer for hyperspectral imaging requires us to spectrally resolve the interferogram given by Eq. (3). The total signal count--rate $R_s$ can thus be spectrally binned into wavelength--dependent counts $R_s(\lambda_s)$. Our setup uses a scanning mirror in the idler arm, which translates parallelly to the idler beam to vary $\varphi_i$ and $\Delta t$. The spectrally resolved count--rate at a mirror position $z$ is thus expected to be of the form
\begin{equation}
    R_s = R_0(\lambda_s)\left[1 + E(z)\mathcal{V}(\lambda_s)\cos\left(\frac{4\pi z}{\lambda_i}+\phi(\lambda_s)\right)\right]
\end{equation}
where $R_0(\lambda_s)$ is the signal count rate outside the coherence range at a signal wavelength $\lambda_s$, $E(z)$ represents the coherence envelope, and $\mathcal{V}(\lambda_s)$ is the maximum interferogram visibility at wavelength $\lambda_s$ over all idler mirror positions. Comparing Eq. (3) and Eq. (5), we conclude
\begin{equation}
\mathcal{V}(\lambda_s) = \frac{R_\text{max}(\lambda_s)-R_\text{min}(\lambda_s)}{R_\text{max}(\lambda_s)+R_\text{min}(\lambda_s)}=|\tau(\lambda_i)|^2
\end{equation}

where $R_\text{min}(\lambda_s)$ and $R_\text{max}(\lambda_s)$ are the wavelength--dependent measured minimum and maximum count--rates in the signal interferogram respectively \cite{Yang:21}. Thus, Eq. (6) enables us to map the object's optical transmission at a conjugate idler wavelength $\lambda_i$ to the corresponding signal photon visibility at $\lambda_s$.

\subsection{Quasi--Phase Matching in PPLN}

The signal--idler photon pairs of interest are generated via Type--0 Spontaneous Parametric Down--Conversion (SPDC) in the PPLN crystal, which is achieved through quasi--phase matching (QPM) between the interacting optical fields. Lithium Niobate possesses a large second-order nonlinear susceptibility, and periodic inversion of its ferroelectric domains reverses the sign of the $\chi^{(2)}$ coefficient. The resulting nonlinear grating introduces an additional reciprocal-lattice wave vector $k_G = 2\pi/\Lambda$, where $\Lambda$ is the poling period. Ignoring temperature--dependence, the QPM condition is exactly satisfied at central signal and idler photon frequencies of $\omega_s^0$ and $\omega_i^0$, given by

\begin{equation}
    \Delta k_{\mathrm{QPM}}(\omega_s^0, \omega_i^0) = k_p - k_s(\omega_s^0) - k_i(\omega_i^0) -\frac{2\pi m}{\Lambda} = 0
\end{equation}
where $k_p,k_s$ and $k_i$ denote the frequency--dependent pump, signal and idler photon wave--vectors respectively and $m$ denotes the QPM order. Moreover, energy--conservation mandates $\lambda_p^{-1} = \lambda_s^{-1} +\lambda_i^{-1}$, where $\lambda_p, \lambda_s$ and $\lambda_i$ are the pump, signal and idler wavelengths respectively. In general, there exists a finite spectral spread in both, signal and idler spectra, reflected in the generated bi--photon state 
\begin{equation}
    |\Psi\rangle_\text{SPDC} = \iint d\omega_1d\omega_2\ \xi(\omega_1,\omega_2)\hat{a}_s^\dagger(\omega_1) \hat{a}_i^\dagger(\omega_2)\ |0\rangle
\end{equation}
where $\hat{a}_s^\dagger(\omega), \hat{a}_i^\dagger (\omega)$ are the creation operators for monochromatic signal and idler photon states with frequency $\omega$ respectively, $|0\rangle$ represents the vacuum state and $\xi(\omega_1, \omega_2)$ is the joint bi--photon spectral amplitude described as \cite{fuenzalida2024nonlinear}
\begin{equation}
    |\xi(\omega_1, \omega_2)| \propto \operatorname{sinc}\left(\frac{\Delta k_\text{QPM}(\omega_1, \omega_2)L}{2} + \mathcal{F}(T)\right)
\end{equation}
where $L$ is the length of the PPLN crystal parallel to the pump beam and $\mathcal{F}(T)$ is the temperature--dependent collinear phase mismatch parameter. The QPM wave--vector mismatch also has an inherent temperature--dependence which originates from the refractive index of the crystal. Consequently, the finite crystal length determines the spectral bandwidth of the generated photon pairs, while the QPM period (set by the crystal channel) and temperature determine the central SPDC wavelengths.

This enables a wavelength--tunable architecture, where the pump beam can be made to pass through different SPDC channels by laterally shifting the crystal, and the temperature can be tuned using an oven. 

\section{Experimental Setup}
A 532 nm continuous wave laser with a narrow bandwidth of $<$ 1 MHz, coupled with a single--mode fiber (NA $= 0.22$) is used as a source, with an output of 64 mW incident on a half--wave plate (HWP), which is necessary for aligning the pump beam polarization with the SPDC crystal specification in order to maximize the SPDC efficiency $\eta$. The laser then falls on a long--pass dichroic mirror (DM--1) which allows the wavelengths above 650 nm to pass and reflects the rest. A 10 cm focal length lens (L1) is used to focus the beam onto the PPLN crystal, which sits in an oven. The crystal, along with the encasing oven is mounted on a mechanical stage, which enables lateral movement for accessing different SPDC channels (each with a different value of $\Lambda$) that are pre--cut into the crystal. Each channel has a length $L = 10\ \text{mm}$ and a cross section of $1\ \text{mm} \times 1\ \text{mm}$.

The collimated pump beam is Gaussian with an FWHM $\sim 0.7\ \text{mm}$. This makes it necessary to focus the beam onto the crystal in order for it to completely fit into the desired channel. Every channel can access a wide range of signal--idler wavelengths that can be tuned using the oven temperature. For our purpose, we use a temperature of \( 110^\circ\mathrm{C} \) to generate a signal--idler pair of specified central wavelengths 810 nm and 1550 nm respectively.

The first signal--idler photons along with the pump are then collimated by lens L2, which is kept at a distance of 10 cm from the crystal center. DM--2, which is a long--pass dichroic mirror at 950 nm allows the idler photons to pass through, while reflecting the signal and the pump. The idler then passes through the sample to be probed, which is mounted on a double--axis motorized stage setup for raster--scanning, transverse to the beam. A scanning mirror $M_i$ is installed on another motorized stage (hereby referred to as the idler mirror), which reflects the idler back along the same path. The signal and pump photons also retrace their respective paths back to the crystal from a fixed signal mirror $M_s$. The pump then does a second--pass through the crystal, generating a second pair of signal and idler photons. As aforementioned, the spatial modes of the idler photons generated from the first and second passes through the crystal are matched by achieving an optimal optical alignment, thus inducing coherence in the signal photons. The pump is then almost entirely reflected back by DM--1, while the signal and idler photons pass through. The residual pump beam and the idler photons are then blocked by an 810 nm band pass filter (BPF{; centre wavelength 810~nm, FWHM 10~nm}), and the signal photons are focused by a 5 cm focal length lens. The detector is placed slightly away from the focal point, where the signal beam is de--focused.

We primarily collect signal--counts data using two detectors: an sCMOS camera, and a CCD spectrometer. The former records cumulative signal counts across a range of wavelengths, thus giving a cumulative interferogram visibility which is lower than the spectrally resolved peak visibility. The latter enables {spectrally resolved} data--collection, {sampling the spectrum at 0.119~nm per pixel. (Section IV A). Details of the optical alignment are given in Appendix B.}

\textbf{Zebrafish sample preparation:} Wildtype zebrafish larvae at 4 days post fertilization (dpf) were fixed in 100\% prechilled acetone for 2 hours at room temperature. The larvae were washed using phosphate-buffer saline to remove any residual acetone traces. Individual larvae were mounted in 70\% glycerol to preserve tissue morphology and optical clearance. The sample was placed within a hole-reinforcement ring on the OHP plastic sheet, and covered with a layer of same to seal the chamber.

\begin{figure}[h]
    \centering
    \includegraphics[width=0.7\columnwidth]{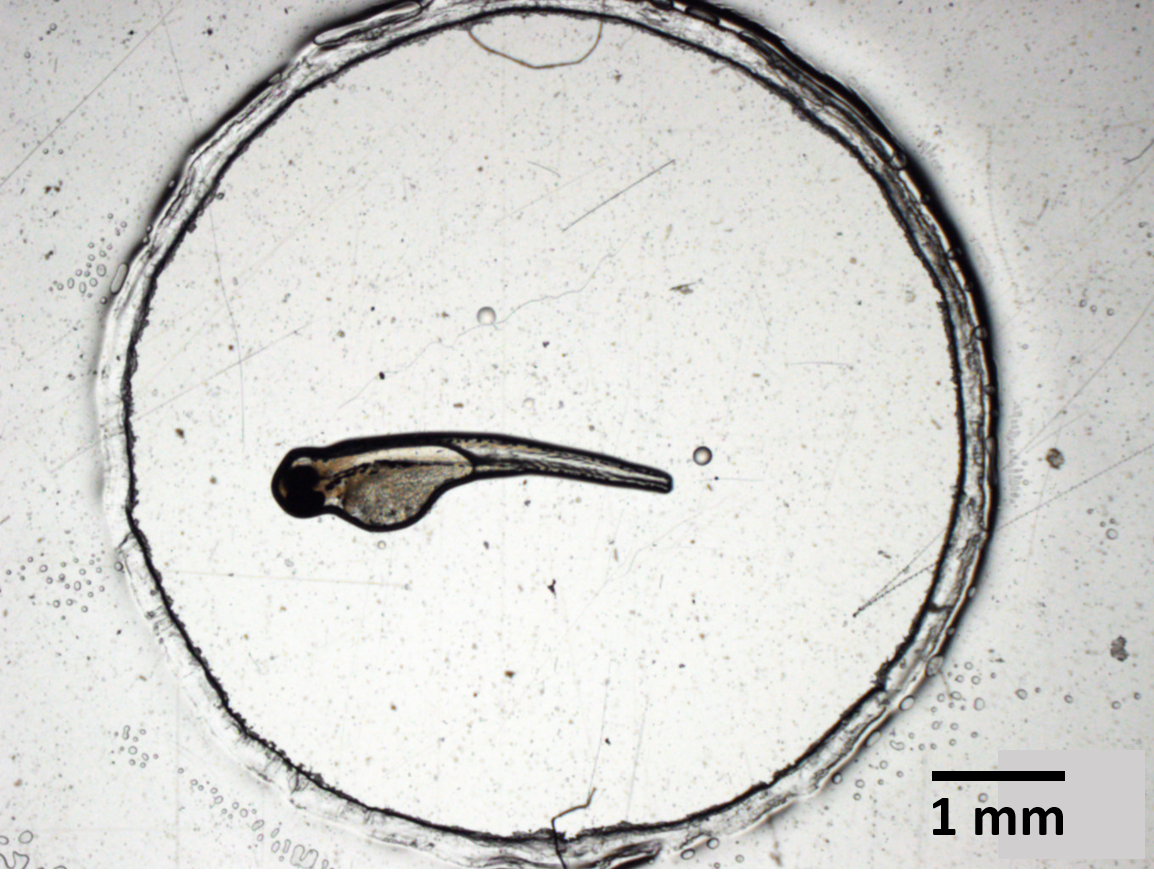}
    \caption{%
        Zebrafish larva sample placed inside a hole--reinforcement ring on an OHP plastic sheet for imaging (Recorded hyperspectral images shown in Figure \ref{fig:results-b})
    }
    \label{fig:zebrafish}
\end{figure}

\section{Results and discussion}
\label{sec:specres}
We assemble the interferometer as given in section III. It is then calibrated by measuring its spectrally resolved baseline visibility $\mathcal{V}(\lambda_s)$. Following this, we measure the spectrally resolved position--dependent optical transmission of our two samples near 1550 nm: (a) zebrafish larva, and (b) a laser--printed carbon ink `LOQM' logo on an OHP plastic sheet, using visibility--based sensing. Both of these results are presented in the following subsections.
\subsection{Interferometer Calibration}
\label{subsec:a}
\begin{figure*}
    \centering
    \includegraphics[width=\textwidth]{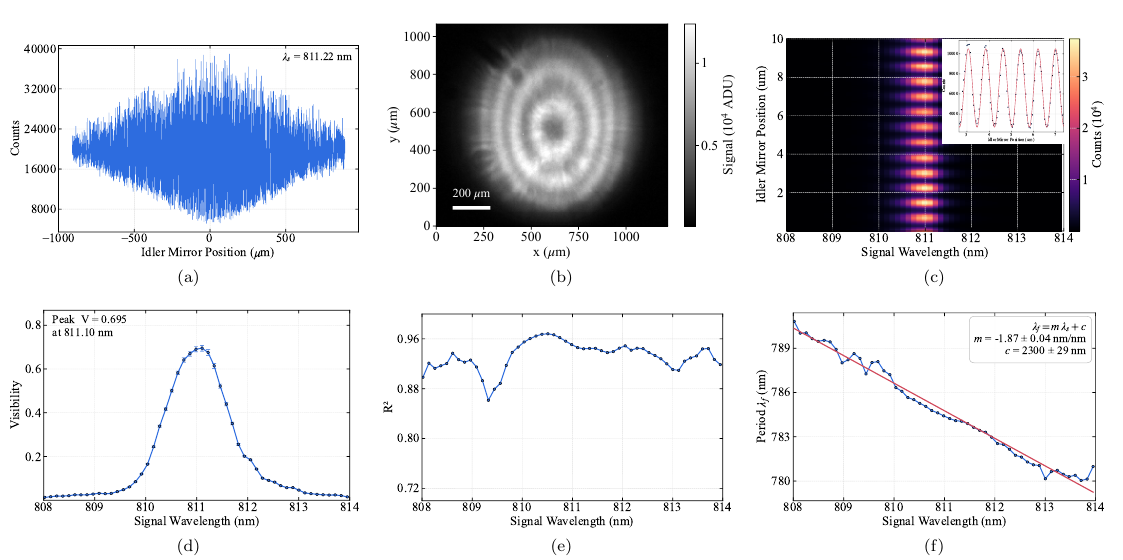}
    
    \caption{Raw interferometer calibration without the sample: (a) Fine scan interferogram with step--size $\Delta z = 1\ \mu\text{m}$ at signal wavelength 811.22 nm; envelope FWHM 1.0~mm, giving the instrument resolution of Section IV A, (b) Spatial interference fringes captured on an sCMOS camera, (c) 2D interferogram showing spectrally resolved interference fringes (inset: Cosine curve fit on raw data at $\lambda_s = 810.51\ \text{nm}$), (d) Baseline visibility $\mathcal{V}(\lambda_s)$ as a function of signal wavelength (without background--subtraction), (e) $R^2$ value for a cosine curve fit on the raw interferogram as a function of wavelength, and (f) Fringe period as a function of wavelength with linear fit (red).}
    \label{fig:composite}
\end{figure*}

As aforementioned, we begin by collecting raw interferometric data without the presence of a sample in the idler arm. Figure \ref{fig:composite} shows the baseline interferometric results. Figure \ref{fig:composite}(a), (b), and (c) show the obtained interferogram in three domains: Coherence envelope at $\lambda_s = 811.22\ \text{nm}$ (obtained by translating the idler mirror stage), spatial planar interference fringes (taken on an sCMOS camera), and spectrally resolved fringes respectively. Figure \ref{fig:composite}(a) reveals a triangular coherence envelope, suggesting a $\text{sinc}^2$ type spectral amplitude of the signal photons. Here, we define the coherence length $L_\text{coh}$ to be the complete range of idler mirror positions over which interference is observed.

{Theoretically, the coherence length can be extracted using the signal--idler group--index difference $\Delta n = 0.079$, obtained by solving the extraordinary--index Sellmeier equation for $\text{MgO:LiNbO}_3$ at 110~$\degree$C, which yields a value $L_\text{coh} = 0.8$~mm \cite{gayer2008} (see Appendix A). A sinc$^2$ fit to the measured signal spectrum gives a true experimental coherence length, $L_\text{coh}^\text{exp} \approx 1$~mm, against 0.8~mm from the Sellmeier equation. The single--channel envelope in Figure \ref{fig:composite}(a) is longer, with a full base $\Delta_\text{ch} \approx 2$~mm. This is not a coherence length of the SPDC light: rather, it is set by the spectral response of one spectrometer channel (the spectrometer's point spread function (PSF)), and binning $N$ adjacent pixels shortens it (as shown in Appendix A), confirming that it is instrument--limited. An sCMOS camera was also used along with an iris in between the BPF and DM--1 to extract the coherence envelope. The interference fringes thus obtained were recorded on a single pixel, which yielded a single--angle coherence length of $\sim 1.5$~mm.} The spectrally resolved baseline visibility is shown in Figure \ref{fig:composite}(d) without dark--counts subtraction, revealing a peak visibility of 69.5\% at 811.1 nm.

Dark--count subtraction gives unreliable results for visibility outside the 810 - 812 nm wavelength window due to noise--domination of the counts. Dark--counts subtraction on the other hand in the 810 - 812 nm window yields a peak visibility of $74.4\%$ at 811.22 nm. {By definition, visibility falls when background counts overtake the SPDC counts. We therefore define the usable spectral window (although a bigger window can be used for samples with a high coefficient contrast) by $\mathcal{V}(\lambda_s)\geq 0.05$, i.e.\ $\lambda_s = 809.6$--813.0~nm ($\lambda_i = 1539.4$--1551.6~nm), an instantaneous coverage of 51~cm$^{-1}$. Across this window, the fringe period follows the energy--conservation prediction (Figure \ref{fig:composite}(f)) and cosine fits retain $R^2 > 0.85$ (Figure \ref{fig:composite}(e)), so the residual fringes carry channel--specific spectral information rather than leakage from the central wavelengths.}
 
The peak visibility is primarily limited by the use of a single lens L2 (refer to Figure \ref{fig:wide_with_inset}) for focusing both: the pump and idler beams back onto the crystal for the second pass. Solving the Sellmeier equations for the central signal--idler wavelengths for an N--BK7 focusing lens yields a focal length mismatch of $\sim 3$ mm, significant enough to cause a mismatch between the idler modes generated upon the first and second passes. A trivial solution is the installation of two lenses: one in the signal and the other in the idler arm after the dichroic mirror DM--2, which can be translated along the direction of propagation to account for the variable focal lengths in both arms, as implemented in \cite{chakraborty2025magneto}. For our experimental setup, L2 was placed on a mechanical stage in order to tune the distance between L1 and L2, for visibility optimization.
 
Figure 3f shows the fringe period $\lambda_f$ as a function of signal wavelength, computed using a cosine curve fit on each wavelength trace (raw data), along with its linear fit. Theoretically,  $\lambda_f = \lambda_i/2$, accounting for the reflection from the mirror $M_i$, which yields the following equation using energy conservation: 
\begin{equation}
\lambda_f (\lambda_s) = \frac{1}{2}\left(\frac{1}{\lambda_p} - \frac{1}{\lambda_s}\right)^{-1}
\end{equation}
This is expected to give the following linear fit at $\lambda_s = 810\ \text{nm}$: $\lambda_f(\text{nm}) \approx 2258\ \text{nm} - 1.83\lambda_s(\text{nm})$. The linear fit obtained experimentally is given in Figure \ref{fig:composite}(f), the equation of which is given by $\hat{\lambda_f}(\text{nm}) = 2300 \pm 29\  \text{nm} - (1.87\pm 0.04)\lambda_s(\text{nm})$, where $\hat{\lambda_f}$ is the least squares linear fit for the fringe period. Both, the $R^2$ value and fringe period against signal wavelengths are in good agreement with the expected trends despite low visibilities outside 810 nm--812 nm.
 
\begin{figure*}[t]
    \centering
 
    \subfloat[]{%
        \includegraphics[width=\textwidth]{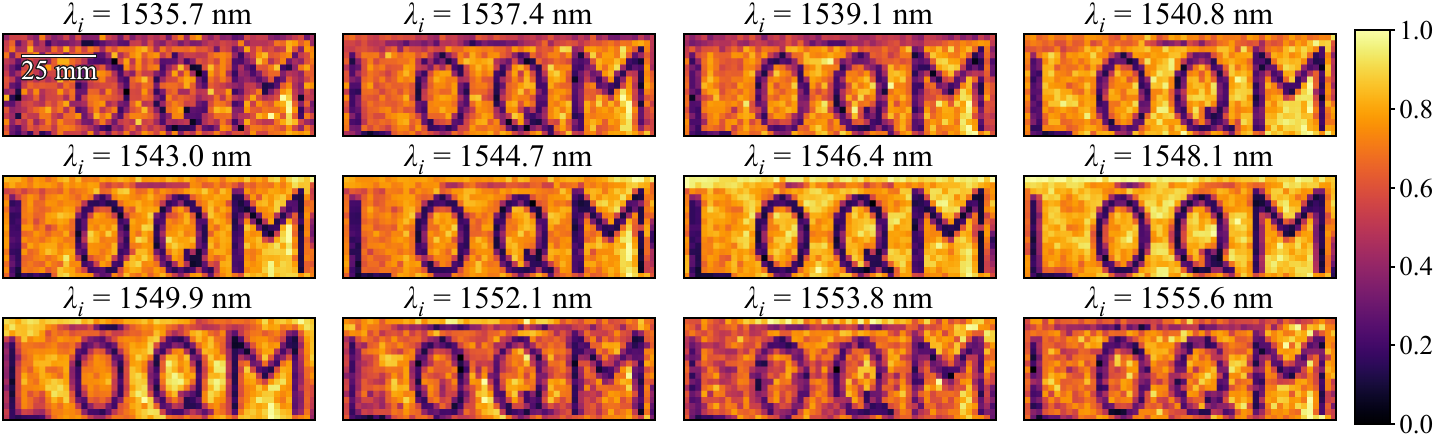}%
        \label{fig:results-a}
    }
 
    \vspace{0.5em}
 
    \subfloat[]{%
        \includegraphics[width=\textwidth]{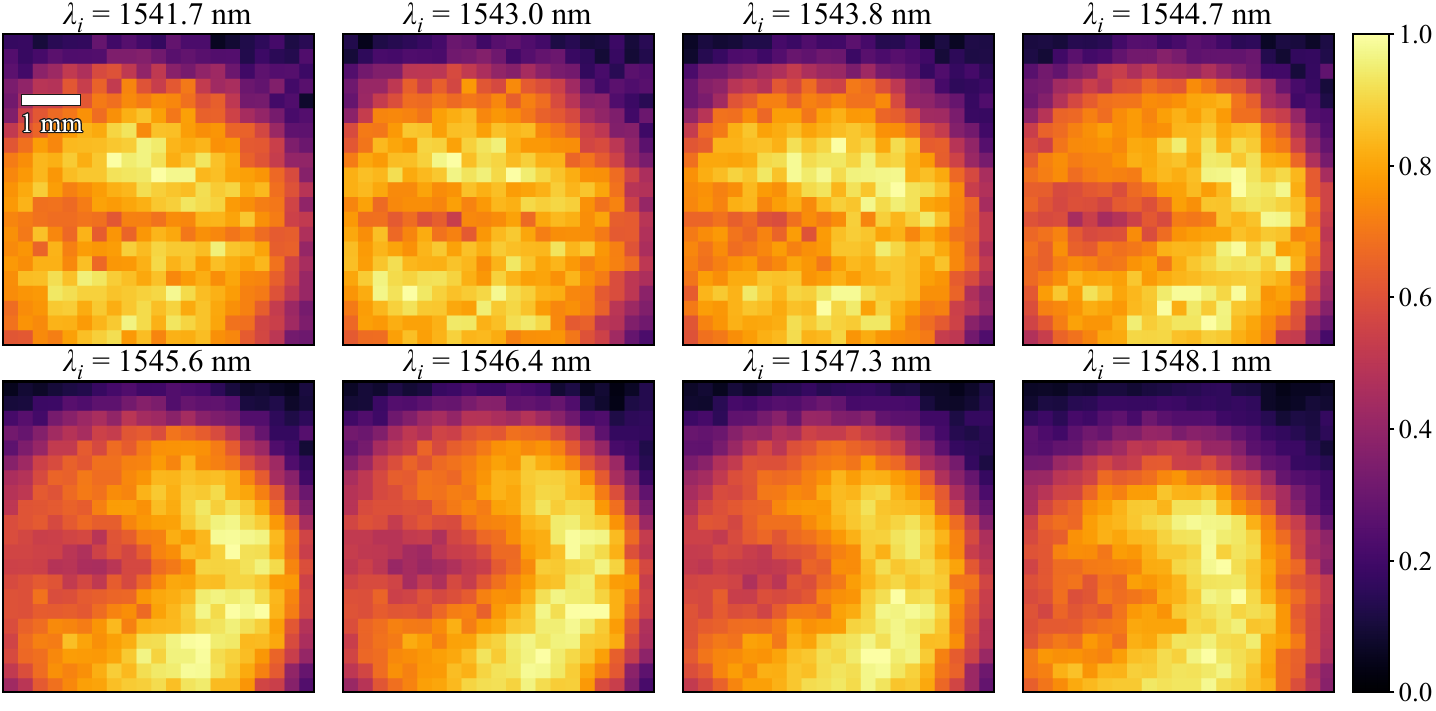}%
        \label{fig:results-b}
    }
 
    \caption{ (a) {Spectrally resolved} visibility--based quantum imaging of an `LOQM' sample (laser--printed on OHP plastic) at specified representative idler wavelengths. The common colour--scale shows the ratios of the detected pixel visibilities in each image, with the measured baseline substrate visibility at the specified idler wavelength. (b) {Spectrally resolved} visibility--based optical response of a zebrafish larva sample (as given in Figure \ref{fig:zebrafish}) at specified representative idler wavelengths. (Note: These images are taken in the sub--resolution limit)}
    \label{fig:results}
\end{figure*}

\textbf{\color{black}Spectral Resolution}\color{black}: With a continuous--wave pump of linewidth below 1~MHz, energy conservation fixes $\omega_i = \omega_p - \omega_s$, so a spectrometer channel of width $\Delta\tilde{\nu}_s$ selects idler photons within
\begin{equation}
\Delta\tilde{\nu}_i = \sqrt{\Delta\tilde{\nu}_s^2 + \Delta\tilde{\nu}_p^2} \approx \Delta\tilde{\nu}_s ,
\end{equation}
where the pump term contributes negligibly at a 1550~nm idler wavelength. The idler--side resolution is therefore that of the signal spectrometer, transferred without loss. Fourier--transform readout does not require a narrow pump \cite{kaur2024pulsed}, but dispersive readout does, and the condition is met here by a factor of $\sim 10^5$.
 
We compute the FWHM of the signal wavenumber $\Delta \tilde{\nu}_s$ from the coherence length. For a triangular coherence envelope, the spectral FWHM in wavenumber follows from the coherence length (full--base of the triangular envelope), as
\begin{equation}
\Delta\tilde{\nu}_s \approx \frac{0.886}{L_\text{coh}}\approx \Delta \tilde{\nu}_i
\end{equation}
With the true experimental coherence length taken to be $L_\text{coh}^\text{exp} \approx 1 $~mm, the SPDC FWHM is given by $\Delta\tilde{\nu}_\text{SPDC} = 8.9\pm 0.13$ (fit) ~cm$^{-1}$. This corresponds to $\Delta \lambda_s \approx 0.58$~nm at the signal, about 5 pixels at 0.119~nm per pixel, and $\Delta \lambda_i \approx 2.12$~nm for the idler, where $\Delta \tilde{\nu} = \Delta\lambda_s/\lambda_s^2 = \Delta \lambda_i/\lambda_i^2$. This yields a resolving power of $R = 730$, defined by $R = \lambda_i/\Delta \lambda_i = 1.129L_\text{coh}/\lambda_i$. The coherence length obtained from the spectrometer's PSF (Figure \ref{fig:composite}(a)) gives an even higher value of $R = 1460$, with $\Delta \tilde{\nu}_i = 4.4\ \text{cm}^{-1}$. Within the 51~cm$^{-1}$ usable window this gives about 12 independent spectral elements, out of which about 2 fall within the experimental SPDC FWHM. Moreover, the measured spectrometer resolution for this single--pixel coherence length is $\Delta \lambda_s = 0.29\ \text{nm}$.
 
Table~\ref{tab:imaging} compares spectrally resolved imaging with undetected photons. Fixed--delay spectral interferometry reaches 8~cm$^{-1}$ \cite{gattinger2026scanless}, pixel--wise Fourier--transform spectroscopy 10~cm$^{-1}$ \cite{placke2025midirhyperspectralimagingundetected}, and filter or phase--matching--defined bands 43--85~nm \cite{paterova2020hyperspectral,kviatkovskyRamelow}; the 4.4~cm$^{-1}$ reported here is comparable to the 4~cm$^{-1}$ commonly used in benchtop FTIR imaging \cite{hughes2015ftir}. The difference follows from the readout. Stepping the idler phase near zero delay demodulates each spectrometer channel independently, so the recovered resolution is the instrument response itself. In fixed--delay schemes the spectrum is carried on a spectral fringe whose sampling by the spectrometer is Nyquist--limited; Kaufmann \textit{et al.} obtained 12~cm$^{-1}$ this way, a factor of 8 below their spectrometer resolution, and 1.5~cm$^{-1}$ with phase scanning \cite{kaufmann2022grating}. Single--point Fourier--transform readout achieves finer resolution, 0.5~cm$^{-1}$ (Table~\ref{tab:pointspec}) \cite{kaur2024pulsed}, but its resolution is set by the delay range: matching 4.4~cm$^{-1}$ requires an optical path difference of $\sim 2.3$~mm, about 2900 Nyquist samples per pixel at 1550~nm, against 114 phase steps here.
\color{black}
\begin{table*}[t]

\color{black}\footnotesize\setlength{\tabcolsep}{4pt}
\caption{Spectrally resolved imaging with undetected photons. $\Delta\tilde{\nu}$: spectral resolution; $R$: idler resolving power; spectral elements: coverage divided by resolution.}
\label{tab:imaging}
\begin{ruledtabular}
\begin{tabular}{lllcccc}
Work & Spectral readout & Idler band & $\Delta\tilde{\nu}$ (cm$^{-1}$) & $R$ & Spatial res. & Elements \\
\hline
This work (spectrometer PSF) & Phase--stepped grating, raster & 1539--1552~nm & 4.4 & $\approx 1460$ & $\approx 1$~mm & $\approx 12^{a}$ \\
This work (true coherence length) & Spectrum & 1539--1552~nm & 8.9 & $\approx $ 730 & $\approx 1$~mm & $\approx 6$\\
Gattinger \cite{gattinger2026scanless} & Fixed--delay grating, raster & 3.3--4.2~$\mu$m & 8 & $\approx 340$ & 12.3~$\mu$m & $\approx 78$ \\
Placke \cite{placke2025midirhyperspectralimagingundetected} & Pixel--wise FTS, wide field & 3.2--4.3~$\mu$m & 10 & $\approx 270$ & $\approx 220$~$\mu$m & $\approx 80$ \\
Paterova \cite{paterova2020hyperspectral} & Phase--scanned camera, tuned bands & 2.75--3.35~$\mu$m & $\approx 46$ & $\approx 70$ & 50~$\mu$m & $\approx 14$ \\
Kviatkovsky \cite{kviatkovskyRamelow} & Tunable filter, wide field & 3.4--4.3~$\mu$m & $\approx 56$ & $\approx 45$ & 35~$\mu$m & $\approx 11$ \\
 \vspace{-4mm}
\end{tabular}
\end{ruledtabular}
\begin{flushleft}\footnotesize $^{a}$Within the $\mathcal{V}\geq 0.05$ window (51~cm$^{-1}$); $\approx 2$ within the SPDC FWHM (8.9~cm$^{-1}$).\end{flushleft}
\end{table*}
 
\begin{table}[t]

\color{black}\footnotesize\setlength{\tabcolsep}{3pt}
\caption{Single--point spectroscopy with undetected photons, and a classical imaging reference.}
\label{tab:pointspec}
\begin{ruledtabular}
\begin{tabular}{llcc}
Work & Readout & Idler band & $\Delta\tilde{\nu}$ (cm$^{-1}$) \\
\hline
Kaur \cite{kaur2024pulsed} & QFTIR, pulsed & 1.53~$\mu$m & 0.5 \\
Lindner \cite{Lindner_2020} & FTS & 3.2--3.9~$\mu$m & 6 \\
Kaufmann \cite{kaufmann2022grating} & Grating, phase scan & 3.2--4.4~$\mu$m & 1.5 \\
Kaufmann \cite{kaufmann2022grating} & Grating, fixed delay & 3.2--4.4~$\mu$m & 12 \\
Gattinger \cite{gattinger2025qftir} & QFTIR & 3.3--4.2~$\mu$m & 4 \\
Hughes \cite{hughes2015ftir} & FTIR imaging$^{a}$ & mid--IR & 4 \\
\end{tabular}
\end{ruledtabular}
\begin{flushleft}\footnotesize $^{a}$Classical benchtop focal--plane--array FTIR imaging, for reference.\end{flushleft}
\end{table}
 
\subsection{Imaging with Undetected Photons}
Figure \ref{fig:results} shows the {spectrally resolved} visibility--based optical response of two samples: \ref{fig:results-b}: a carbon ink laser--printed logo on an OHP plastic sheet, and \ref{fig:results-a}: zebrafish larva (sub--resolution imaging). Each image is reconstructed using the signal interferogram at a specific signal wavelength, the corresponding idler wavelength of which, is computed using energy conservation. Each image is given a normalized colour--grading to enable a direct comparison in contrast across wavelengths.
 
The limiting values on the colour--scales are the ratios of the maximum and minimum visibilities computed over the entire sample with the baseline substrate visibility at the specified idler wavelength $\tilde{\mathcal{V}}(\lambda_i)$. The baseline substrate visibility is the mean visibility of the raw substrate (mounting material without the sample). This is computed by averaging the pixel visibilities over 5 pixels in each hyperspectral image which have the highest recorded visibilities. The modulus of the transmission coefficient at a pixel $(x,y)$ of an image taken at an idler wavelength $\lambda_i$ is given by
\begin{equation}
    |\tau(\lambda_i,x,y)|^2 = \frac{\mathcal{V}(\lambda_i, x,y)}{\mathcal{V}(\lambda_s)}
\end{equation}
where $\mathcal{V}(\lambda_i,x,y)$ denotes the computed visibility at an idler wavelength $\lambda_i$ at pixel $(x,y)$, and $\mathcal{V}(\lambda_s)$ is the baseline visibility at the corresponding signal wavelength (Figure \ref{fig:composite}(d)). Thus, the normalized visibilities used in Figure \ref{fig:results} represent the modulus squared of the relative transmission coefficient at each pixel of every image with respect to the substrate (single OHP plastic sheet in the case of the carbon ink target, and a double OHP plastic sheet in the case of the zebrafish larva), given by
\begin{equation}
    \mathcal{V}_\text{norm}(\lambda_i,x,y) = \Bigg|\frac{\tau(\lambda_i,x,y)}{\tau_\text{substrate}(\lambda_i)}\Bigg|^2 = \frac{\mathcal{V}(\lambda_i,x,y)}{\tilde{\mathcal{V}}(\lambda_i)}
\end{equation}
The data was taken on a spectrometer {sampled at 0.119~nm per pixel, with a measured resolution of $\Delta \lambda_s = 0.29\ \text{nm}$ (see section IV A).} Both samples were 2D raster--scanned: The zebrafish larva with a resolution $\Delta x = \Delta y = 250\ \mu\text{m}$, and the LOQM logo with a resolution $\Delta x = \Delta y = 2\ \text{mm}$. The zebrafish larva was imaged over $21 \times 21$ pixels, while the logo over $52 \times 17$ pixels. For the zebrafish, the visibility at each pixel was calculated using signal counts taken over an idler mirror scanning range of $8\ \mu\text{m}$ with a step--size of 70 nm, while for the logo, a range and step--size of $4\ \mu\text{m}$ and 70 nm respectively were used. The zebrafish images were recorded over a duration of $\sim 8$ hours, while that of the logo were recorded over a duration of $\sim 12$ hours. A spatial drift of the order of $\sim 2$ mm was observed in higher idler wavelengths above the central idler wavelength, as described in detail in Appendix D. Details of the visibility calculation algorithm used are given in Appendix C.
 
\section{Conclusion}
Through this paper, we successfully demonstrate the ability of our wavelength--tunable experimental setup to record {spectrally resolved} images in the near--IR and IR regimes{, at a measured spectral resolution of 4.4~cm$^{-1}$, the finest reported for imaging with undetected photons. Two upgrades follow directly: temperature tuning of the quasi--phase--matching condition extends the coverage across 1100--2700~nm at high resolutions, and a higher--dispersion spectrometer raises the resolution, which remains $\sim 10^5$ above the pump--linewidth limit}. At present, the spatial resolution of our images is limited by the idler beam waist ($\sim 1\ \text{mm}$ FWHM), which is the main hindrance to microscopic imaging with enhanced resolution. This can be fixed by installing a $4f$ lens system in the idler arm, and placing the sample to be imaged at the focal point. This shall lead to a significant reduction in the idler beam spot size on the sample, enabling a microscopic resolution power. 
 
Further, the developed experimental setup has the potential to be used for infrared absorption spectroscopy of gases such as carbon dioxide and water vapour, which show significant absorption in the near--IR and IR regimes. The setup can also be compacted for industrial use in Fourier--Transform Infrared Spectroscopy (FTIR). 
\begin{acknowledgments}
We acknowledge funding support from the National Quantum Mission, an initiative of the Department of Science and Technology, Government of India. A.K. acknowledges the support from ANRF via grant number CRG/2022/001170.
\end{acknowledgments}

\bibliography{References}
\newpage 
\appendix 
\section{Coherence Envelope}
Consider two identical photon wave--packets, each described by the normalised 1D $z$--space wavefunction $A(z)$. Let their separation along the $z$ direction be $z_0$. Thus, the total photonic amplitude is given by
\begin{equation}
    \psi(z) = A(z) + A(z-z_0)e^{-ik_0z_0}
\end{equation}
where $k_0$ is the effective wave--vector for the propagation of these photons. The normalised average photon count--rate $R$  is thus given by 
\begin{equation}
\int_{-\infty}^{\infty}|\psi(z)|^2\ dz = \int_{-\infty}^\infty |A(z) + A(z-z_0)e^{-ik_0z_0}|^2\ dz
\end{equation}
which can be simplified to the following form:
\begin{equation}
R = 2  + 2E(z_0)\cos(k_0z_0)
\end{equation}
where $E(z_0)$ denotes the coherence envelope given by
\begin{equation}
E(z_0) = \int_{-\infty}^\infty A^{*}(z)A(z_0-z)\ dz = (A^{*}*A)(z_0)
\end{equation}
assuming $A(z) = A(-z)$. The $k$--space wavefunction of each photon is given by $\tilde{A}(k) = \mathcal{F}^{-1}\left[A(z)\right]$ where $\mathcal{F}^{-1}$ denotes the inverse Fourier transform operation. Note that the IFT of $E(z)$ is given by the convolution theorem
\begin{equation}
\mathcal{F}^{-1}\left[E(z)\right] = |\tilde{A}(k)|^2 = I(k)
\end{equation}
where $I(k)$ is the normalised $k$--space intensity distribution of each photon. Alternatively, Fourier--transforming both sides of Eq. (A5) gives the $k$--space analogue of Eq. (4):
\begin{equation}
    E(z) = \int_{-\infty}^\infty I(k)e^{ikz}\ dk
\end{equation}
The above equation can be trivially transformed into frequency--space to yield Eq. (4). {For a continuous--wave pump, pairs are created uniformly along the crystal, and signal and idler
walk off at a rate set by the group-index mismatch $\Delta n = n_s-n_i$.
The single-photon amplitude is therefore an estimate of width equal to the walk-off length,
\begin{equation}
A(z) = \frac{1}{\sqrt{\ell}}\,\mathrm{rect}\!\left(\frac{z}{\ell}\right),
\qquad \ell = \Delta n L .
\end{equation}
where $L$ is the crystal length. Its Fourier transform gives the familiar phase-matching spectrum,
\begin{equation}
I(k) \propto \mathrm{sinc}^2\!\left(\frac{\Delta k L}{2}\right)
     = \mathrm{sinc}^2\!\left[\frac{\Delta n L\,(k-k_0)}{2}\right],
\end{equation}
so the sinc of phase matching and the finite crystal length are conjugate descriptions. The interference envelope is the autocorrelation of $A(z)$,
equivalently the Fourier transform of $I(k)$, which is a triangle:
\begin{equation}
E(z_0) = \Lambda\!\left(\frac{2z_0}{\ell}\right) =
\begin{cases}
1 - 2|z_0|/\ell, & |z_0| \le \ell/2 \\
0, & |z_0| > \ell/2
\end{cases}
\end{equation}
The envelope has straight sides and terminates sharply, with no tails. Its full base defines
the coherence length, $L_\mathrm{coh} = \ell$.
Using the half-maximum of $\mathrm{sinc}^2$, the spectral and spatial widths are related by
\begin{equation}
\Delta\tilde{\nu}\, L_\mathrm{coh} = 0.886,
\qquad
L_\mathrm{coh} = 0.886\,\frac{\lambda_i^2}{\Delta\lambda_i},
\end{equation}
so the usual estimate $\lambda^2/\Delta\lambda$ overstates the coherence length by about 13\%.
All lengths are mirror--travel distances, which have to be doubled for their corresponding optical delay counterparts.} 

\begin{figure}[h]
    \centering
    \includegraphics[width=1\columnwidth]{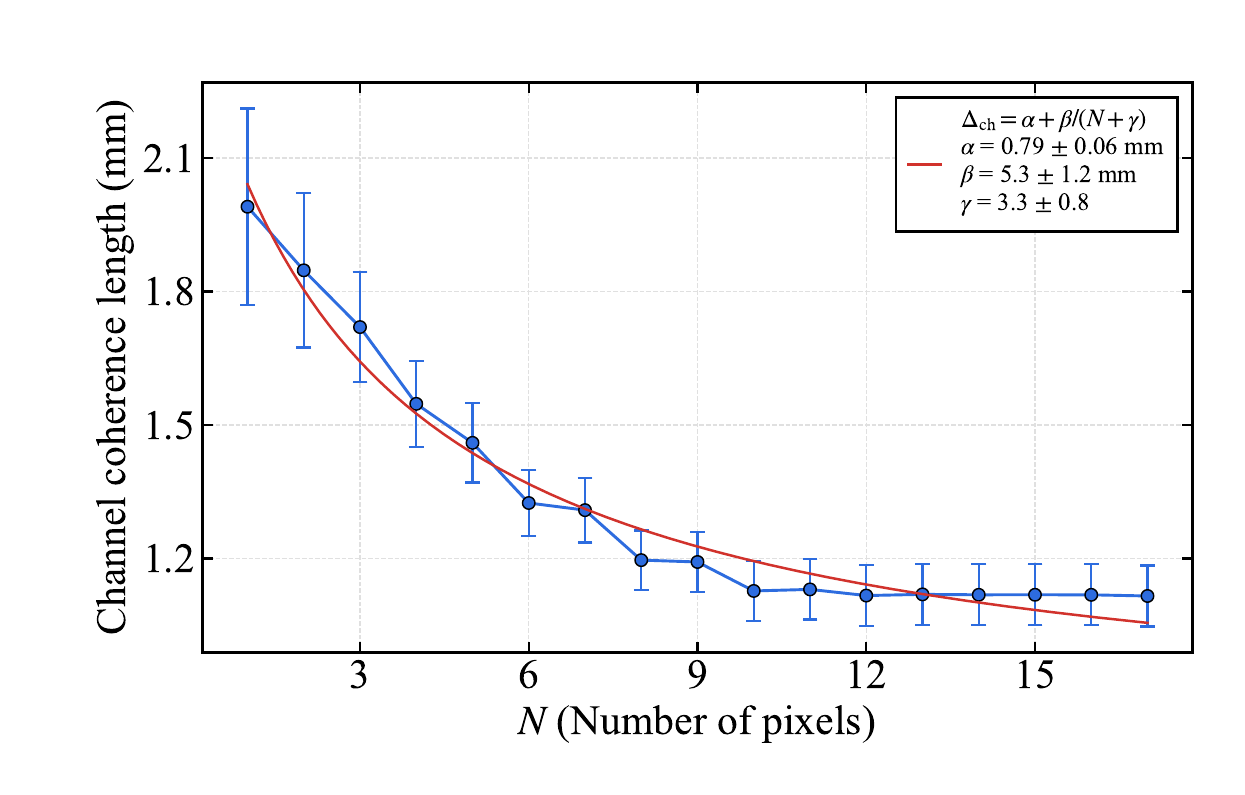}
    \caption{%
        Recorded channel coherence length $\Delta_\text{ch}$ plotted against the number of spectrometer pixels $N$ in the channel
    }
    \label{fig:pixels}
\end{figure}

The channel--dependent coherence length $\Delta_\text{ch}$ as reported in Figure \ref{fig:pixels} was found by varying the number of spectral bins $N$ considered (centered around $\lambda_s = 811.22\ \text{nm}$) in the channel. For $N = 1$, only a single spectrometer pixel was considered (whose envelope is given in Figure \ref{fig:composite}(a)). For an odd $N >1$, the counts in $(N-1)/2$ neighbouring pixels on either side of the central one were summed over, while for an even $N>1$, $N/2$ and $(N-2)/2$ pixels were considered to the left and right of the central one respectively. The value of $\Delta_\text{ch}$ saturates to roughly $1.11\ \text{mm}$, against our previously quoted experimental coherence length of $1$ mm, owing to uncertainty in the envelope fitting (for full--base length extraction). A hyperbolic fit of the data is as shown in red in Figure \ref{fig:pixels}. As aforementioned, each pixel encompasses a wavelength bin $\delta \lambda_s = 0.119\ \text{nm}$. Thus, the total wavelength range contained in $N$ pixels is $\Delta \lambda_s(\text{nm}) = 0.119N$. Substituting for $N$ in the hyperbolic fit, we obtain
\begin{equation}
    \Delta_{\text{ch}} = (0.79 \pm 0.06) + \frac{5.3 \pm 1.2}{(\Delta \lambda_s/0.119)+(3.3\pm 0.8)}
\end{equation}
where $\Delta_\text{ch}$ and $\Delta\lambda_s$ are taken to be in mm and nm respectively, and $\Delta\lambda_s$ is the spectral width of the channel centered around $\lambda_s = 811.22\ \text{nm}$. 
\section{Optical Alignment}
Any nonlinear interferometric setup that leverages Induced Coherence requires a precise alignment of all optical components involved to ensure a good degree of spatial mode--matching between the idler photons generated from the two SPDC processes. In the following section, we describe a precise standard operating procedure for the alignment of a folded nonlinear interferometer:
 
We hereafter refer back to Figure \ref{fig:wide_with_inset} for the description of the experimental setup. We start the setup alignment by using a driving mirror (not shown here), along with DM--1, in order to drive the pump laser along a path exactly parallel to a row of holes on the optical table (this shall become the main arm of the interferometer).The idler mirror, along with its translating motorized stage is then affixed onto the table. The tilt--knobs of the idler mirror are adjusted so that the beam spots from the laser, and after reflection from the idler mirror on both, DM--1 and the driving mirror are perfectly overlapping. For a more precise alignment, an sCMOS camera is mounted at the position of the detector, and a focusing lens of focal length 5 cm is placed in between the camera and DM--1. The motorized stage is then translated back and forth to check for any drift in the residual pump beam leak on the camera. Upon observing a slight drift, the idler mirror tilt--knobs are re--adjusted until it is eliminated. The sCMOS camera is then centered with this pump leak as the reference, using a GUI generated reticle.
 
The second dichroic mirror DM--2 is then installed, roughly at $45$ degrees to the interferometer arm, followed by the installation of the stationary signal mirror $M_s$. The tilt--knobs on $M_s$ are adjusted so that the residual reflected pump beam is aligned with the reticle on the previously fixed sCMOS camera. 
 
\begin{figure}[h]
    \centering
    \includegraphics[width=1\columnwidth]{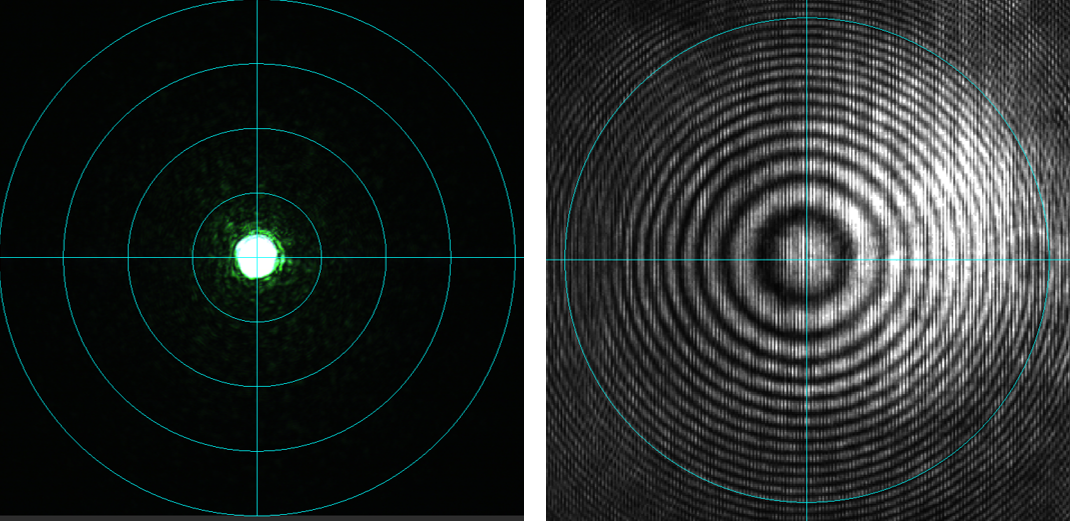}
    \caption{%
        Alignment of L1 using its focused beam spot on the CMOS camera (left), and back--reflection on the sCMOS camera (right)
    }
    \label{fig:alignment}
\end{figure}
 
The first lens L1 is then aligned as follows: Before placing the lens, a CMOS camera is placed roughly at the Fourier plane and is centered with reference to the collimated pump beam. This is followed by the lens installation. The translational degrees of freedom of the lens are tuned by ensuring that the focused beam spot is centered on the CMOS camera, while the tilt degrees of freedom are tuned by centering the back--reflection of the lens (which looks like an airy--pattern) on the sCMOS camera (shown in Figure \ref{fig:alignment}).
 
This is followed by removing the CMOS camera and installing the second lens L2 roughly at a distance of 20 cm from L1, whose translational degrees of freedom are tuned by centering the reflected collimated beam spot on the sCMOS camera, and the tilt degrees of freedom, by centering the back--reflection, similar to the process followed for L1. Finally, the oven--encased crystal is then mounted on a three--axis mechanical stage, placed roughly equidistant from both lenses, and laterally shifted to access the desired SPDC channel. At last, the BPF is installed in front of the collecting lens, followed by the installation of an iris between DM--1 and L1 for eliminating background signal counts, significantly enhancing the visibility.
 
\section{Visibility Calculation}
 
The images shown in Figure \ref{fig:results} are recorded with an idler mirror stage step--size of 70 nm, the fringe period being $\sim 775$ nm, thus yielding about 11 points per fringe. Despite satisfying the Nyquist criterion, detector noise and jitter from the scanning stage leads to somewhat noisy data, with an unreliable cosine curve fit. Thus, an accurate calculation of fringe visibility demands the use of an alternate method that does not rely on fitting a cosine curve on the raw data. The following section describes the methodology used for the same.
 
\begin{figure}[h]
    \centering
    \includegraphics[width=1\columnwidth]{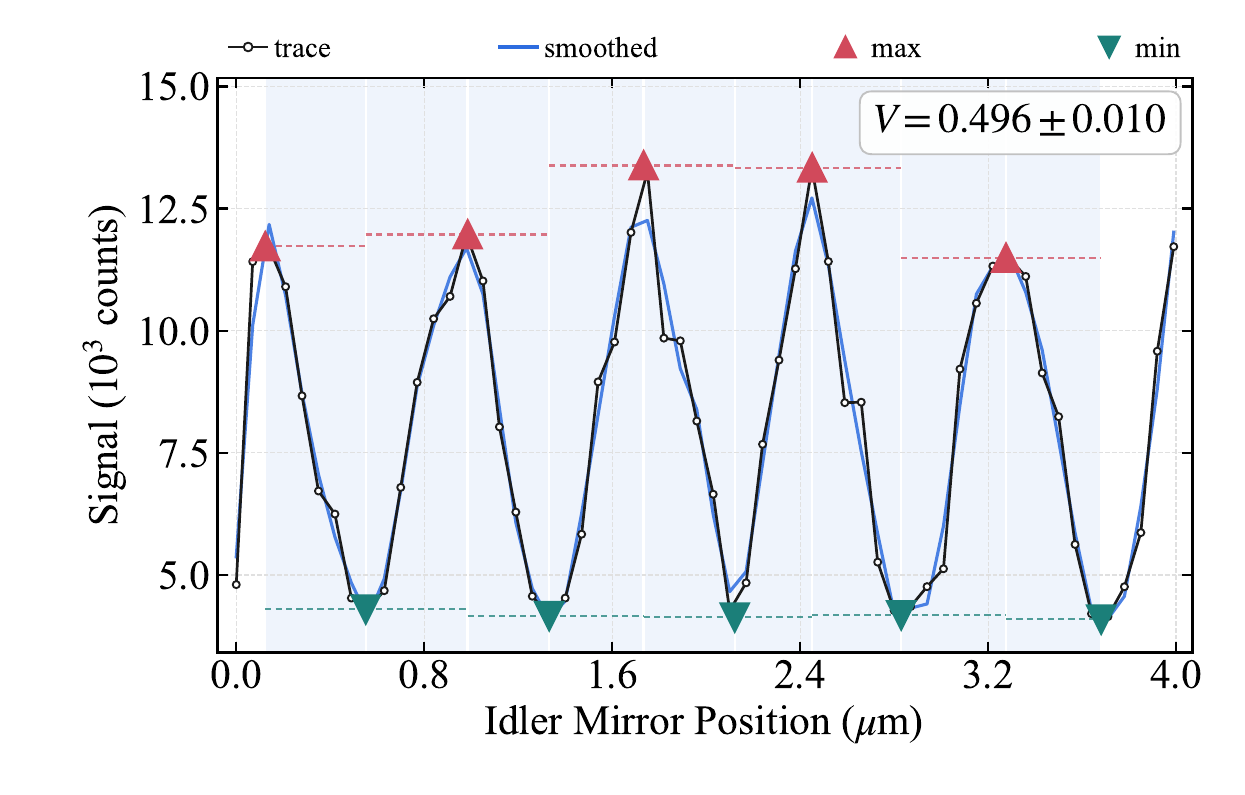}
    \caption{A single pixel trace illustrating peak and trough--detection for computing local visibilities
    }
    \label{fig:peak}
\end{figure}

The spectrometer's exposure time is set to 200 ms, and for every measurement of signal counts at a particular pixel (sample position) and idler mirror position, a set of 10 measurements is taken, and its average $I_k$ and standard deviation $\sigma_k$ are computed. The dark counts are then subtracted, giving a data--point $y_k = I_k - d_k$ where $d_k$ denotes the averaged dark counts taken at a wavelength far away from 810 nm. For each pixel trace, the mean noise is calculated as 
\begin{equation}
    \sigma^{(i)} = \frac{1}{n}\sum_{k=1}^{n}\sigma_k^{(i)}
\end{equation}
This is then multiplied by a prominence factor to obtain the prominence floor, which is fed into a Python code for peak--detection. The prominence floor sets a threshold for the amount by which a peak value must exceed the local minima values on either side for it to be considered a valid local maximum. The peak--detection algorithm is then run on data smoothed using a Savitzky Golay filter, $\tilde{y}_k$. The algorithm is applied on $\tilde{y}$ for peak--detection, and on $-\tilde{y}$ for trough--detection. This is followed by a sub--sample refinement of the local extrema detected. For each extremum at index $i$, a parabola through the three raw points $(z_{i-1}, y_{i-1}),\ (z_i, y_i)$ and $(z_{i+1},y_{i+1})$ is solved in closed form. With $h = (z_{i-1} + z_{i+1})/2$, $D = y_{i-1} - 2y_i + y_{i+1}$ and $\delta = (y_{i-1} - y_{i+1})/2D$, we get 
\begin{equation}
z^{*} = z_i + \delta h,\ \ y^{*} = y_i - \frac{1}{4}\left(y_{i-1} - y_{i+1}\right)
\end{equation}
The refinement is rejected and the raw extremum is kept if $D = 0$, $h\leq 0$, $|\delta|>1$, or if $z^{*}$ falls outside $[z_{i-1}, z_{i+1}]$. A working example of the peak detection algorithm for a pixel trace is shown in Figure \ref{fig:peak}.
 
For an accepted maximum--minimum pair (indexed $j$), where $y^* = N_\text{max}^{(j)}$ at the maximum and $y^* = N_\text{min}^{(j)}$ at the minimum, we calculate the local visibility $\mathcal{V}_j$ as
\begin{equation}
    \mathcal{V}_j = \frac{N_\text{max}^{(j)} - N_\text{min}^{(j)}}{N_\text{max}^{(j)} + N_\text{min}^{(j)}}
\end{equation}
Propagating independent errors through Eq. (C3) with $\partial \mathcal{V}_j/\partial N_\text{max}^{(j)} = 2N_\text{min}^{(j)}/(N_\text{max}^{(j)} + N_\text{min}^{(j)})^2$ and $\partial \mathcal{V}_j/\partial N_\text{max}^{(j)} = -2N_\text{max}^{(j)}/(N_\text{max}^{(j)} + N_\text{min}^{(j)})^2$, we get
\begin{equation}
    \sigma_{\mathcal{V}}^{(j)} = \frac{2}{(N_\text{max}^{(j)} + N_\text{min}^{(j)})^2}\sqrt{N_\text{max}^{(j)2} \sigma_\text{max}^{(j)2}+ N_\text{min}^{(j)2}\sigma_\text{min}^{(j)2}}
\end{equation}
where $\sigma_\text{max}$ and $\sigma_\text{min}$ denote the computed standard deviations of the $y^*$ values at both the extrema from the individual standard deviations at $y_{i-1},\ y_i$ and $y_{i+1}$ (notations pertain to sub--sample refinement). These local visibilities are then averaged over the trace of each pixel to obtain the final spectrally resolved pixel visibility. The error--bars shown in Figure \ref{fig:composite}
(d) are also computed using Eq. (C4). In some cases, the peak--detection algorithm fails to identify any prominent peaks and troughs in the trace data, for which interpolation with surrounding pixel visibilities is used to compute the visibility.
 
\section{Angular dependence and Spatial shift}
 
\begin{figure}[h]
    \centering
    \includegraphics[width=1\columnwidth]{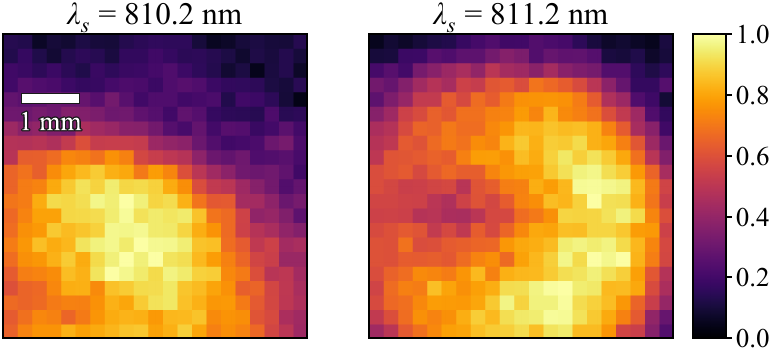}
    \caption{%
        A spatial shift of $\sim 2$ mm observed between two zebrafish images: At the central signal wavelength (right), and at a $\sim 1$ nm blue--shifted wavelength (left) (Refer to Section IV B for colour--scale details)
    }
    \label{fig:shift}
\end{figure}

Figure \ref{fig:shift} shows an observed spatial shift of $\sim 2$ mm in the recorded zebrafish sample images across a wavelength difference of $\Delta \lambda_s \approx -1\ \text{nm}$ from the central signal wavelength $\lambda_s^{0} \approx 811.1\ \text{nm}$, at which the maximum visibility is measured. This shift is only observed for $\Delta \lambda_s < 0$, indicative of a finite angle of dispersion of the idler photons from the PPLN crystal for $\lambda_i > \lambda_i^{0}$, where $\lambda_i^{0}$ is the central idler wavelength. 
 
In the following section, we give a quantitative explanation for this phenomenon. Let $q$ be the transverse wave--vector for some wavelength $\lambda$ when exiting the crystal at an angle $\theta$. Then, we trivially have $\sin \theta = q\lambda/2\pi$. Now, the longitudinal component $k_z$ of the wave--vector can be approximated as
\begin{equation}
k_z = k\sqrt{1 - \frac{q^2}{k^2}} \approx k - \frac{q^2}{2k}
\end{equation}
Thus, $\Delta k_z$ inside the crystal can be written in terms of $\Delta k = k_p - k_s - k_i - 2\pi m/\Lambda$ (see Eq. (7)) for a signal frequency $\omega_s$ as
\begin{equation}
    \Delta k_z = \Delta k(\omega_s) + \frac{q^2}{2}\left(\frac{1}{k_s} + \frac{1}{k_i}\right)
\end{equation}
by applying Eq. (D1) to both signal and idler photons. Now, $\Delta k(\omega_s)$ can be Taylor expanded about the central signal frequency $\omega_s^0$. Defining $\Omega = \omega_s - \omega_s^0$, we get
\begin{equation}
\Delta k(\omega_s) \approx \Delta k(\omega_s^0) + \frac{d\Delta k}{d\omega_s}\Bigg|_{\omega_s^0}\Omega = \frac{d\Delta k}{d\omega_s}\Bigg|_{\omega_s^0}\Omega
\end{equation}
since by the QPM condition, $\Delta k(\omega_s^0) = 0$. Further,
\begin{equation}
    \frac{d\Delta k}{d\omega_s} = -\frac{dk_i}{d\omega_s} - \frac{dk_i}{d\omega_i}\frac{d\omega_i}{d\omega_s} = -\frac{n_s - n_i}{c}
\end{equation}
where $c/n_s$ and $c/n_i$ are the signal and idler group velocities in the crystal respectively. Thus, to a first order approximation in $\Omega$, $\Delta k(\Omega) = -\Delta n\Omega/c$. Substituting this into Eq. (D2) and setting $\Delta k_z$ to zero gives us the following value of $q$ which satisfies the QPM condition:
\begin{equation}
    q_\text{QPM}^2(\Omega) = \frac{2k_sk_i}{k_s + k_i}\frac{\Delta n}{c} \Omega 
\end{equation}
We note that Eq. (D6) will have a solution only if $\Omega > 0$, which implies $\lambda_s < \lambda_s^0$, in agreement with Figures \ref{fig:results-a} and \ref{fig:shift}. Thus, for $\lambda_s > \lambda_s^0$, the $q$ value satisfying the QPM condition is $q = 0$, giving a null spatial shift, while $\lambda_s < \lambda_s^0$ yields a finite spatial shift. 
 
Converting to the wavelength space and using $d\omega/d\lambda = -2\pi c/\lambda^2$, we get $\Omega = -2\pi c \Delta \lambda_s/(\lambda_s^0)^2$. Substituting for $\Omega$ in Eq. (D6) and solving for the angle of dispersion in the small angle regime, we obtain an idler dispersion angle given by
\begin{equation}
    \theta_i = \frac{\lambda_i^0}{\lambda_s^0}\sqrt{-\frac{\bar{k}\Delta n \Delta\lambda_s}{\pi}} \sim \sqrt{\lambda_s^0 - \lambda_s}
\end{equation}
where $\bar{k} ^{-1} = k_s^{-1} + k_i^{-1}$. The value of $\Delta n = n_s - n_i$ for PPLN can be computed using the Sellmeier equation for the extraordinary index $n_e$ of MgO--doped Lithium Niobate at $T = 110 \degree$C \cite{gayer2008}. This yields $\Delta n = 0.079$, which upon substituting in Eq. (D7) gives $\theta_i \approx 23$ mrad. 
 
For small dispersion angles, the spatial shift is given by $\delta = D\theta$ where $D$ is the distance between the crystal and lens L2. For our case, $D = 100\ \text{mm}$. Thus, for a signal wavelength blueshift of $\sim 1\ \text{nm}$, the expected spatial shift in the formed visibility--based image is expected to be $\approx 2.3\ \text{mm}$, against an experimentally observed value of $\sim 2\ \text{mm}$ as shown in Figure \ref{fig:shift}. This spatial shift can be avoided by the use of a 4f lens system in the idler arm, with the sample placed on the focal plane.
 \\
\section{Mean Photon Number per Mode}

The following section provides the calculation for an upper bound of the mean photon number $\mu$ per electromagnetic mode in our nonlinear interferometer.
 
For a continuous--wave pump, a mode is defined by a combination of a single transverse mode, coherence time and polarization. Type--0 phase matching ensures a single polarization state. The mean photon number per mode is thus given by 
\begin{equation}
    \mu = \frac{\Phi_\text{gen}\tau_c}{M}
\end{equation}
where $\Phi_\text{gen}$ is the generated photon flux, $M$ is the number of collected transverse modes and $\tau_c$ is the coherence time defined as $\tau_c = \lambda_s^2/c\Delta \lambda_s$, where $\Delta \lambda_s$ is the collected signal bandwidth. The generated photon flux can be written in terms of the detected count rate $R_\text{det}$ as $\Phi_\text{gen} = R_\text{det}/\eta$ where $\eta$ is the total detection efficiency including collection, optics, fiber coupling, and detector quantum efficiency. An upper bound on $\mu$ can be calculated by setting $M = 1$. Considering {the measured SPDC bandwidth of 0.58~nm}, a detected count rate of $\sim 10^5\ \text{s}^{-1}$ (order of magnitude estimate) and a total efficiency of $10\%$, we obtain 
\begin{equation}
    \color{black}\mu < 3.7 \times 10^{-6} \ll 1
\end{equation}
The above bound being much smaller than unity {places the source deep in the spontaneous regime, excluding} stimulated emission as the source of the interference fringes.
\end{document}